\documentclass[12pt]{iopart}

\newcommand{\ket}[1]{\left|#1\right\rangle}

\newcommand{\braket}[2]{\left \langle #1 |  #2\right\rangle}
\newcommand{\ketbra}[2]{\left| #1 \rangle \langle #2 \right|}
\newcommand{\Ca}{Ca$^{+}~$}
\newcommand{\avg}[1]{\langle #1 \rangle}

\usepackage[utf8x]{inputenc}
\usepackage{graphicx}
\usepackage{dcolumn}
\usepackage{bm}
\usepackage{subfigure}
\usepackage{subfloat}

\begin{document}

\title{Trapped ions in optical lattices for probing oscillator chain models}

\author{Thaned Pruttivarasin, Michael Ramm, Ishan Talukdar, Axel Kreuter, Hartmut H\"affner}

\address{Department of Physics, University of California,  Berkeley, CA 94720, USA}

\ead{hhaeffner@berkeley.edu}

\begin{abstract}

We show that a chain of trapped ions embedded in microtraps generated by an optical lattice can be used to study oscillator models related to dry friction and energy transport. Numerical calculations with realistic experimental parameters demonstrate that both static and dynamic properties of the ion chain change significantly as the optical lattice power is varied. Finally, we lay out an experimental scheme to use the spin degree of freedom to probe the phase space structure and quantum critical behavior of the ion chain.

\end{abstract}

\maketitle

\section{Introduction}

Quantum simulators explore the emergence of phenomena by reconstructing systems of interest with well-controlled building blocks such as trapped atoms or ions. The simulations provide the means to analyse and understand the physics of large and computationally intractable quantum systems \cite{Feynman1982,Lloyd1996}. As a consequence, there is a lot of excitement in building such quantum simulators motivated further by the prospects of studying novel quantum phases and phase transitions \cite{Bloch2008,Porras2004,Friedenauer2008,Kim2009a}. 

Quantum simulations can also be used to study non-linear quantum systems, quantum chaos and the quantum-to-classical transition. Furthermore, it is an exciting question whether two quantum simulators simulating the same Hamiltonian would agree with each other. Any differences between two such simulations not explained by experimental imperfections hint towards new physics. 

Current efforts in quantum simulations with trapped ions concentrate, mainly, on global interactions among spins \cite{Friedenauer2008,Kim2010,Johanning2009}. While a linear arrangement of the ions is typically considered, Clark \textit{et al.} \cite{Clark2009} also study ions trapped in a two-dimensional grid and placed close to each other to mediate spin-spin interactions. However, Clark \textit{et al.} conclude that such a design is likely to fail because the coupling between ions is too weak as compared to the decoherence rate due to motional heating from nearby electrodes.

In this paper, we consider an array of microtraps formed by an optical lattice superimposed with the harmonic potential of a conventional Paul trap \cite{Schmied2008}. The scheme enables strong coupling without electrodes nearby. The optical lattice can exert considerable forces onto the ions \cite{Schneider2010} with trap frequencies exceeding 1~MHz, while the external potential holds the ions closely together, thus allowing for strong coupling between ions via the Coulomb force. By increasing the lattice strength, the system transitions from a regime where the ion-ion coupling dominates to a regime where the ions form a chain of independent oscillators. Most importantly, such microtrap potentials are anharmonic and, thus, exhibit interesting quantum dynamics.

In the following, we focus on the motional degree of freedom of trapped ions rather than on the spin or electronic degrees of freedom. In section~\ref{sec:frenkel}, we show that trapped ion oscillator chains are well suited to study the emergence of important phenomena such as dry friction. In section~\ref{sec:energy-trans}, we discuss how the oscillator chains can be used to study energy transport \cite{May2007,Rebentrost2009,Riera2011}. The strength of the microtraps can depend on the internal state of the ions. Based on this, we show in section~\ref{sec:spin-motion-dynamics} that an interferometric method can be used to probe the phase space structure and quantum critical behaviours of the ion chain. 

\section{Simulating friction: the Frenkel-Kontorova Model}
\label{sec:frenkel}
\subsection{Introduction}

\begin{figure}
\begin{center}
\includegraphics[width = 0.6\textwidth]{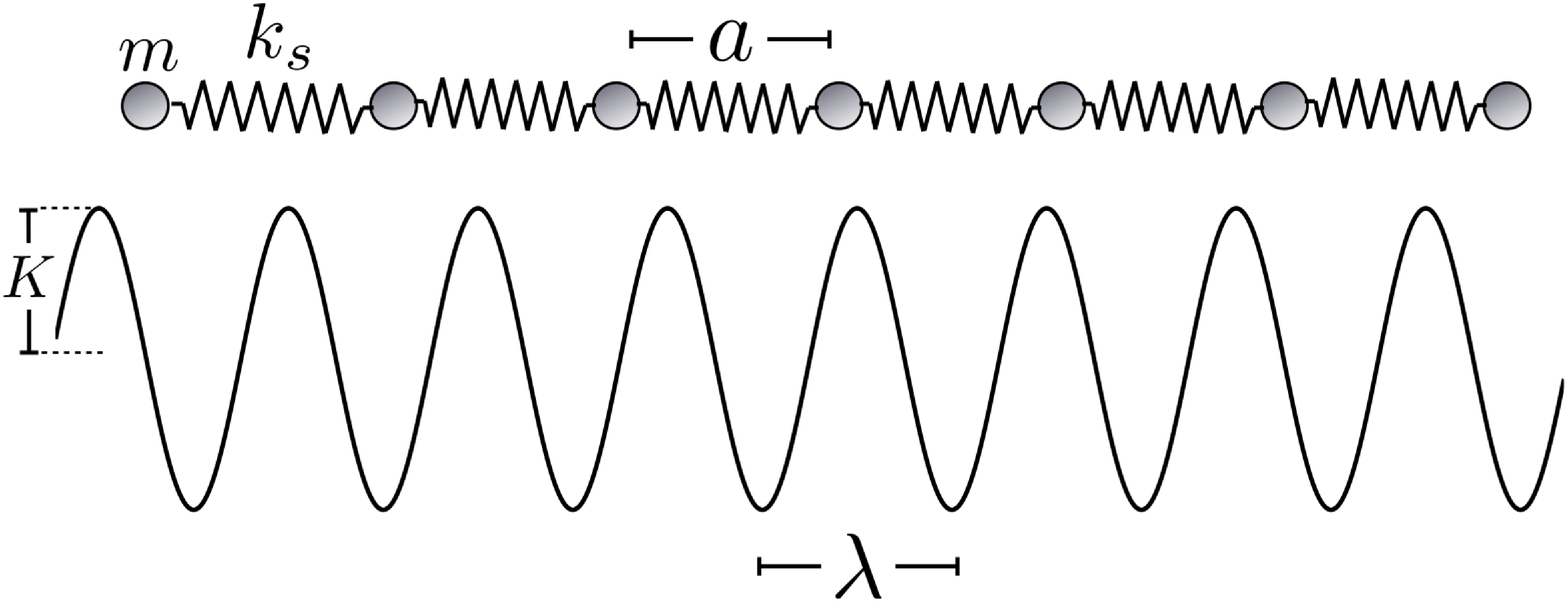}
\caption{\label{FK_model}The Frenkel-Kontorova model. A one-dimensional chain of identical masses connected by identical springs is perturbed by an external periodic potential.}
\end{center}
\end{figure}

The Prandtl-Tomlinson model \cite{Prandtl1928,Tomlinson1929} describes friction as the motion of a point mass along a sinusoidal potential \cite{Popov2010}. While this model explains basic properties of dry friction such as static friction and the onset of kinetic friction, it does not incorporate internal degrees of freedom, which are needed to describe heating. The Frenkel-Kontorova model (FK) \footnote{Also called the Frenkel-Kontorova-Tomlinson model in the study of dry friction.} allows for heating by assuming multiple particles with mass $m$ interconnected via springs with spring constant $k_s$ in a sinusoidal potential of strength $K$ and period $\lambda$ (see figure~\ref{FK_model}). This model yields a more realistic description of friction between two crystals: a chain of atoms from one crystal sliding along the periodic potential created by the other crystal \cite{Frenkel1938,Kinoshita2006,Brauna}.

The Hamiltonian of the FK model is given by
\begin{equation}
H = \sum_{i}^{\infty}\left[\frac{1}{2}m\dot{x_i}^2+\frac{1}{2}k_s(x_i-x_{i-1}-a)^2+K\cos{\left(\frac{2\pi x_i}{\lambda}\right)}\right],
\end{equation}
where $x_i$ is the displacement of the $i$-th particle and $a$ is the natural length of the spring. The parameter $a$ is the lattice constant when $K=0$. From this point on, we assume that the external periodic potential is incommensurate with the lattice constant, i.e., $a/\lambda$ is an irrational number \footnote{The analysis of the commensurate case can be found in \cite{Brauna}.}.

In the context of dry friction, we define the ``depinning force", $F^{\delta x}_d$, as the force required to displace the chain by a small amount $\delta x$. In the limit where $K = 0$, the system has translational symmetry and, hence, $F^{\delta x}_d$ is zero. On the other hand, if $K$ is large, each particle will be pinned to the minimum of the external periodic potential and $F^{\delta x}_d$ is non-zero. Aubry proves that $F^{\delta x}_d$ vanishes for $K$ from zero up to a critical value $K_c$, but when $K>K_c$, the depinning force $F^{\delta x}_d$ becomes finite \cite{Aubry1983,Peyrard1983}. We call the phase where $F^{\delta x}_d = 0$ the sliding phase and where $F^{\delta x}_d \neq 0$ the pinned phase. The transition from the sliding to the pinned phase is usually called the Aubry transition. 

The persistence of the sliding phase for finite but small $K$ stems from the incommensurability between the chain and the periodic potential. Thus, at low $K$, the particles can be found at every phase of the periodic potential and the effect of the periodic potential cancels \cite{Aubry1983,Peyrard1983}. In particular there is no restoring force on the center-of-mass mode of the chain. As $K$ increases beyond $K_c$, particles start to avoid the maxima of the periodic potential and translational symmetry is broken. For the chain to move, at least one of the particles needs to move over a maximum and, thus, a finite $F^{\delta x}_d$ is required. A detailed analysis can be done by transforming the system to an area-preserving map and applying the Kolmogorov-Arnold-Moser theorem, which is discussed extensively in~\cite{Aubry1983,Peyrard1983}.

\subsection{Frenkel-Kontorova Model in trapped ions (FKI)}

A laser cooled ion crystal provides an ideal system for studying the physics connected to the FK model \cite{Garc'ia-Mata2007,Benassi2011}. In the regime where the confinement in the radial direction is significantly stronger than the axial confinement, the ion crystal becomes one-dimensional and the Hamiltonian can be written as 
\begin{equation}
H = \sum_{i}^{}\left[\frac{1}{2}m\dot{x_i}^2+\frac{1}{2}m\omega_a^2x_i^2+K\cos{\left(\frac{2\pi x_i}{\lambda}\right)}\right]+\sum_{i<j}\left[ \frac{e^2}{4\pi\epsilon_0 |x_j-x_i|}\right],
\label{FKH}
\end{equation}
where $\omega_a$ is the axial trap frequency. In contrast to the FK model,  the interaction of the ions with each other is long range and anharmonic. Furthermore, the ion-ion spacing typically is not uniform. However, the ions mimick the incommensurability condition by sampling each phase of the lattice.

Attempting to characterize the system in the same way as the FK model, we find that the depinning force, $F^{\delta x}_d$, is not zero at $K=0$ because of the presence of the axial confinement. Nevertheless, Garc\'{\i}a-Mata \textit{et al.} show that despite this departure from the FK model, there still exists a phase transition between the two phases corresponding to the sliding and the pinned phases of the FK model \cite{Garc'ia-Mata2007}. However, one has to relax the definition of the sliding phase defined in the original FK model by considering the relations between $F^{\delta x}_d$ and $K$. For the pinned phase, $F^{\delta x}_d$ increases monotonically as we increase $K$ whereas in the sliding phase, $F^{\delta x}_d$ does not necessarily increase monotonically with $K$ and can even decrease as $K$ is increased. 

Following Benassi \textit{et al.}, another approach is to characterize the phase transition by imposing that the centre ion is positioned at the maximum of the lattice, which makes the system reflection symmetric about the axial trap centre \cite{Benassi2011}. At large $K$, this symmetry is broken and a finite $\tilde{F}^{\delta x}_d$ is the force required to restore the reflection symmetry of the system. The restoring force $\tilde{F}^{\delta x}_d$ can be used as an order parameter that characterises the structural phase transition.

The analyses in both \cite{Garc'ia-Mata2007} and \cite{Benassi2011} assume that the ratio of the period of the external periodic potential to the ion-ion spacing near the axial trap center is given by the golden ratio. At these large lattice constants of a few $\mu$m, the phase transition occurs at a very high optical lattice intensity requiring excessive laser powers on the order of kilowatts. A shorter period of the optical lattice alleviates this power requirement due to an increased curvature of the lattice potential. Furthermore, the ions need to be displaced less to avoid the maxima of the lattice as compared to longer lattice periods. Combined, both effects lead to a quadratic reduction in the required power. In the following section, we show that the phase transition indeed survives for the lattice period on the order of 200~nm and, thus, the structural phase transition becomes accessible using realistic experimental parameters. 

\begin{figure}
\begin{center}
\includegraphics[width = 0.7\textwidth]{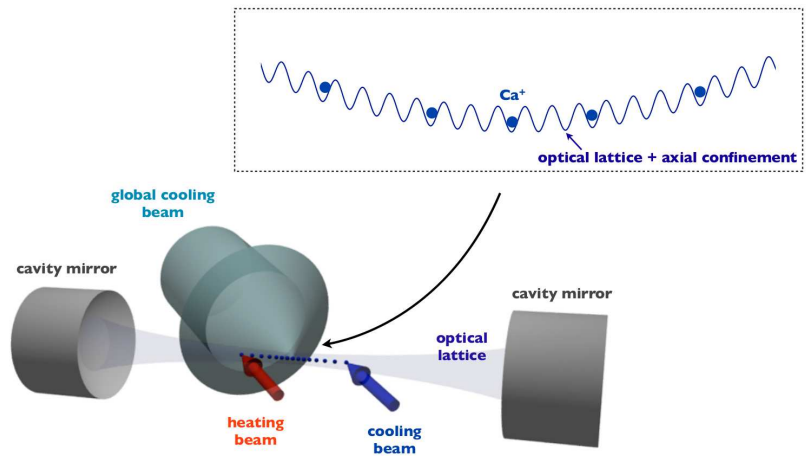}
\caption{\label{setup} A schematic of the proposed setup. A linear ion string is trapped in a conventional Paul trap and placed in an optical lattice between two mirrors along the axial direction. The ion chain is kept sufficiently cold by laser cooling with a beam large enough to cover the whole chain. Two separated tightly focused laser beams can be used to heat, cool, and measure the temperature of individual ions. The inset shows the total axial potential experienced by the ions from the optical lattice and the harmonic confinement from the trap electrodes (not shown in the figure).}
\end{center}
\end{figure}

\begin{figure}
\begin{center}
\includegraphics[width = 0.4\textwidth]{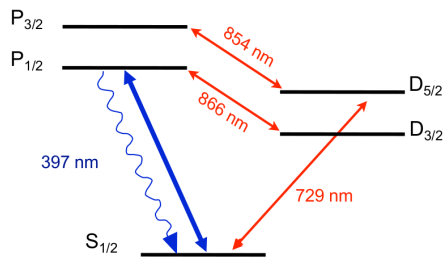}
\caption{\label{energy_level}Energy level of calcium ions. The $S_{1/2}-P_{1/2}$ dipole transition at 397~nm is used for Doppler cooling. The narrow $S_{1/2}-D_{5/2}$ quadrupole transition at 729~nm is used for state detection and measuring the temperature. 866~nm and 854~nm are used to prevent an ion from being in dark states.}
\end{center}
\end{figure}

\subsection{Numerical calculations}

\begin{figure}
\begin{center}
\includegraphics[width = 0.85\textwidth]{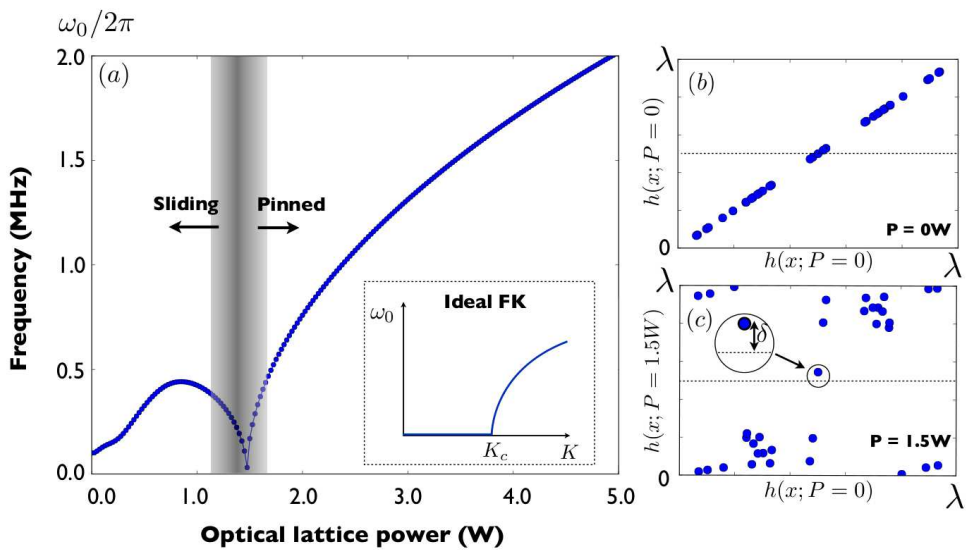}
\caption{\label{omega_K_calculation} (a) The lowest eigenfrequency, $\omega_0$, of a 35-ion chain plotted against the optical lattice power. The inset shows the behaviour of the FK model: below the Aubry transition, $\omega_0=0$. In the main plot, although $\omega_0$ does not start from zero, there is a clear transition point beyond which $\omega_0$ increases monotonically with the lattice power. Recalling the definition of the sliding and pinned phase for the FKI model, we see that there are two regimes. In the sliding phase, $\omega_0$ does not always increase as we increase the lattice laser power. In particular, at a critical point of the laser power, i.e., 1.5~W for 35-ion chain, one of the eigenmode frequencies approaches zero and for this eigenmode the electrostatic forces are exactly balanced by the optical forces. Beyond the critical value, $\omega_0$ increases monotonically with the lattice laser power, and hence the system is in the pinned phase. 
(b,c) Hull functions for an optical lattice power of 0~W and 1.5~W. At P = 1.5~W, all ions avoid the maxima of the optical lattice, which is indicated by the dashed line. $\delta$ is the displacement of the center ion from the lattice maximum.}
\end{center}
\end{figure}

For the calculations, we assume \Ca trapped in a standard Paul trap as shown in figure~\ref{setup}. An optical lattice can be conveniently generated using a diode laser at 405~nm, detuned by 8~nm from the $S_{1/2}-P_{1/2}$ dipole transition of \Ca (see figure~\ref{energy_level}) and coupled into an optical cavity. We assume a cavity waist of 25~$\mu$m corresponding to a Rayleigh range of 5~mm. With an axial confinement $\omega_a = 2\pi \times 100$~kHz, the length of a 35-ion string is 230~$\mu$m which is significantly shorter than the Rayleigh range. The optical lattice strength can then be assumed to be uniform over the ion chain.

The calculations are carried out by first minimizing the total energy of the ion chain to find the equilibrium position of each ion using the full Coulomb coupling. To simulate the experimental procedure, the optical lattice power is then increased in small increments. The new equilibrium positions are calculated by choosing the step size in the minimizing algorithm small enough to prevent any ion from jumping to a neighbouring lattice site. At each lattice power, a coupling matrix $A_{ij} = \partial^2 H/\partial x_i \partial x_j$ is calculated numerically. Assuming that the center ion initially sits at a maximum of the lattice, there is a critical lattice strength where the Coulomb force from the trap and the other ions on the center ion are exactly balanced by the optical force. At this critical lattice strength the lowest eigenfrequency of the matrix $A_{ij}$ vanishes (see figure~\ref{omega_K_calculation}a) and the center-of-mass mode of the ion chain slides freely. For stronger lattices, we enter the pinned phase, while for weaker lattices, we call the system to be in the sliding phase. 

We can make the sliding to pinned phase transition more transparent using the hull function, $h(x;P)$. It measures the displacements of the ions (at power $P$) from the original positions (at $P=0$) modulo the lattice wavelength. Once the system is in the pinned phase, the ions avoid the maxima of the optical lattice indicated by the center dashed line in figure~\ref{omega_K_calculation}b and \ref{omega_K_calculation}c. At a lattice power of 1.5~W, no ion is found at a lattice maximum. The deviation of the center ion from the lattice maximum, $\delta$, shown in figure~\ref{omega_K_calculation}c, can be used as an order parameter analogous to the depinning force, $F_{d}^{\delta x}$, discussed earlier.

\subsection{Experimental considerations}

Assuming the parameters above, around 1.5~W of the optical lattice power is required to observe the phase transition. Coupling about 5~mW of laser power into a cavity with finesse of 500 yields around 2~W of intra-cavity power for the lattice. Consequently, one can tune from the sliding phase to the pinned phase simply by increasing the power of the laser coupled into the cavity. The ion string can be cooled either directly in the lattice or if the AC-Stark effect causes severe complications, by loading an already cooled ion string adiabatically into the the lattice. 

To cool all axial modes of the ion string to the ground state electromagnetically induced transparency (EIT) cooling seems appropriate as multiple modes can be cooled close to the ground state simultaneously by tailoring the absorption profile of the ions such that multiple modes are addressed at the same time \cite{Morigi2000,Roos2000a}. Alternatively, sideband cooling addressing might be employed to reduce the temperature even further \cite{Monroe1995a,Schmidt-Kaler2000,Leibfried2003,Haeffner2008,Morigi2003}. While with EIT cooling simultaneous cooling of two vibrational modes of a single ion has been demonstrated \cite{Roos2000a,Morigi2000}, it is not clear whether sideband cooling can be carried out on multiple modes simultaneously. For sequential sideband cooling, on the other hand, one has to contend with reheating of all the modes that are not actively cooled due to photons scattered during the process. This shortcoming could be partially mitigated by first pre-cooling close to the ground state with EIT cooling to reduce the effective Lamb-Dicke parameter before starting sideband cooling, lowering the probability of the spontaneously emitted photons to recoil. Thus, we expect that EIT cooling followed by sideband cooling will allow for efficient ground state cooling on a reasonable time scale.

The heating from the scattering of the lattice photons (scattering rate of 40 events per watt per second) is suppressed by the Lamb-Dicke parameter $\eta=k_z z_0$. Here $k_z$ is the component of the photon wavevector along the trap axis, and $z_0=\sqrt{\hbar/2m\omega_{\rm{loc}}}$ is the typical size of the ground state wavefunction associated with the local trap frequency of $\omega_{\rm{loc}}$ along the trap axis where we neglect the less important radial direction. The ion will also experience heating from fluctuations in the laser intensity. We identify two sources of heating: parametric heating and heating from  a fluctuating gradient force. When trapped at lattice minimum, parametric heating is dominant. The mean rate of increase in the motional energy is given by $\Gamma_{\rm heat}=\pi^2\nu^2S(2\nu)$, where $\nu$ is the trap frequency generated by the optical lattice \cite{Gehm1998}. With the aid of active intensity stabilization, the relative intensity noise can be maintained to within $S(\nu) \approx 10^{-14} $ Hz$^{-1}$ \cite{Mosk2001}. This yields a heating rate of $10^{-4}$ quanta/ms from the ground state (at $\nu = 1.2$ MHz, associated with the lattice power of 1.5~W).  When the ion is positioned at the maximum slope of the lattice well, intensity fluctuations produce an gradient force. We calculate the transition matrix element of such a perturbation, $H'=\epsilon(t)x(dH_{\rm{lattice}}/dx)$, where $H_{\rm{lattice}}=K\cos{(2\pi x/\lambda)}$ \cite{Gehm1998}.  Here $\epsilon(t)=(I(t)-\avg{I(t)})/\avg{I(t)}$ is the fractional fluctuation in intensity and $K/\hbar=2\pi\times6.9$ MHz is the lattice trap depth at a lattice power of 1.5 W. This yields a heating rate of $2\pi^3 K^2 S(\nu)/\hbar\nu m \lambda^2 = 0.04$~quanta/ms, where $\lambda =202.5~$nm is the lattice wavelength. Such rates are comparable to typical heating rates of macroscopic Paul traps \cite{Leibfried2003,Haeffner2008} and should not be a serious impediment on the experiment time scales of less than $\sim$ 10 ms.

The mode spectrum of the ion chain can be measured either by a photon correlation method \cite{Rotter2008} or via conventional sideband spectroscopy \cite{Leibfried2003,Haeffner2008}. The depinning force can be measured directly by applying differential voltages on the DC electrodes and measuring the displacement of the center ion using a CCD camera or via methods presented below for measuring the hull function.

The hull function of the ion chain can be obtained by measuring the AC-Stark shift of the individual ions in the optical lattice using interferometric techniques \cite{Haeffner2003a}. For this, one can use a narrow auxiliary transition such as the $S_{1/2}$ to the $D_{5/2}$ quadrupole transition at 729~nm (see figure~\ref{energy_level}) to prepare a superposition state. Letting the ions evolve for some time and recombining the two wavefunctions yields the phase evolution of the $S_{1/2}$ with respect to the $D_{5/2}$ state. Since this measurement scheme provides the position modulo the lattice period, combining this method with standard CCD camera imaging yields the position of each ion to a sub-lattice-wavelength resolution.

\subsection{Discussion}

So far we have not considered any quantum mechanical effects. Garc\'{\i}a-Mata \textit{et al.} show that, with finite quantum fluctuations, the system exhibits observable structural phases similar to those distinguished by the classical Aubry transition \cite{Garc'ia-Mata2007}. However, the calculations are carried out with the middle part of the ion chain, where the ion-ion distance is approximately uniform, and only the nearest-neighbour Coulomb interaction is included \cite{Garc'ia-Mata2007}. Although there have been many studies related to quantum effects in the FK model \cite{Ho2000,Hu2006,Zhirov2003,Zhong2006}, a full quantum mechanical treatment for the FKI model is difficult due to the long range nature of the Coulomb force. This underlines the importance of quantum simulators for a full understanding of the FKI model. Another uniquely quantum mechanical effect is tunnelling. With a short optical lattice wavelength, tunnelling is expected to be prominent, especially at the point close to the transition from the sliding to pinned phase. Well-established techniques of ground state cooling mentioned above can be used to reach the quantum mechanical ground state, which is important in probing this tunnelling effect near the transition point.

In this section, we have explored the static properties of the FK system and its realization using trapped ions. Next, we investigate the system from a dynamical point of view, namely, the propagation of excitations along the ion chain.

\section{Energy Transport}
\label{sec:energy-trans}
\subsection{Introduction}

\begin{figure}
\begin{center}
\includegraphics[width = 0.9\textwidth]{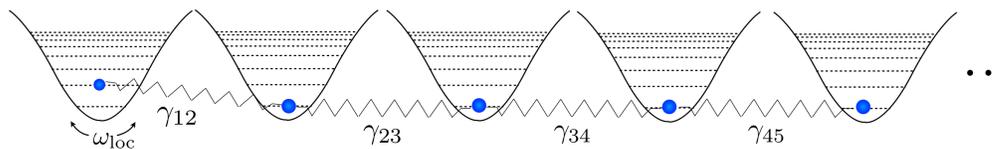}
\caption{\label{fig:coupled_oscillators} Excitation transport along a chain of coupled oscillators. Each particle oscillates in an anharmonic microtrap generated by the optical lattice with an oscillation frequency $\omega_{\rm{loc}}$. Coulomb couplings between the $i$-th and the $j$-th oscillator are characterised by $\gamma_{ij} = \partial^2 H /\partial x_i \partial x_j$. For simplicity, the figure shows only the nearest-neighbour couplings.}
\end{center}
\end{figure}

A string of ions each trapped in a microtrap forms a chain of oscillators coupled to each other via the Coulomb interaction (see figure~\ref{fig:coupled_oscillators}). Such oscillator chain models are extensively used to study quantum transport within molecules \cite{May2007,Rebentrost2009}, for example, light harvesting complexes \cite{Engel2007}.

One important topic from both a technological and fundamental standpoint is the transport of energy and quantum information along the oscillator chains \cite{Plenio2005}. For instance, it is conjectured that there is a certain amount of decoherence which optimises efficiency of energy transport in light harvesting complexes \cite{Caruso2009,Chin2010}. However, it has been shown that highly efficient transport phenomena can also arise within a fully quantum mechanical framework \cite{Scholak2010,Scholak2011}. In naturally occurring systems, the degree and quality of decoherence is hard to control, making it difficult to establish the exact origin of the highly efficient energy transport processes.

In contrast, in the ion oscillator chain, the amount of decoherence can be controlled with laser beams addressing the ions individually. Furthermore, individual addressing can be used to prepare a very diverse set of quantum states of the chain, e.g., phonon excitations, with coherent state manipulation \cite{Leibfried2003,Haeffner2008}. In addition, coupling parameters in the system can be tuned. For instance, the effective coupling strength of the oscillators is varied by changing the axial trapping potential or the optical lattice strength. The numerical calculations in the previous section show that both the strong coupling regime (ion-ion coupling comparable to the individual oscillator frequency as in the sliding phase) and the weak coupling regime (ion-ion coupling much smaller than the oscillator frequency as in the pinned phase) can be attained using realistic experimental parameters.

Previous theoretical studies have revealed interesting properties of ion strings in the absence of the lattice potential related to the inherent long-range ion-ion interactions, non-uniform spatial distributions \cite{Morigi2004,Morigi2004a}, and non-linear effects \cite{Marquet2003}. Here, we focus on heat and energy transport measurements with and without an optical lattice. 

\subsection{Energy and heat transport}

Two approaches can be adopted to study the system: (1) investigating the dynamics by following phonon excitations as they propagate along the ion crystal and, (2) observing how the system equilibrates when brought into contact with thermal baths. In the following, we discuss both types of experiments with long ion chains.

Figure~\ref{setup} shows a schematic of a setup well-suited for such experiments. In order to study energy propagation within the crystal, one can apply a short pulse of a tightly focused resonant laser beam to one end of the chain, locally exciting the ion motion, while monitoring the temperature at the other end of the chain. The ion temperature can be measured by analysing the Doppler profile, for instance, of the dipole $S_{1/2}-P_{1/2}$ transition, or a narrow transition such as the $S_{1/2}-D_{5/2}$ quadrupole transition. An important requirement is that the time resolution of the measurement is higher than the energy transport time. The rate of energy propagation depends on the eigenmode spectrum of the long ion chain, so it is of particular interest to measure the effect of the optical lattice on the propagation time for various lattice powers.

It is necessary to heat the ion quickly compared to the energy transport time. We estimate the attainable rate with which the ion can be heated by assuming that it has first been cooled close to the ground state. Then, we switch on an intense resonant beam with a saturation parameter $s \gg 1$ where $s = 2 \Omega^2/\Gamma^2$ with the on-resonance Rabi frequency of $\Omega$ and the natural linewidth of $\Gamma$. The ion experiences a force $F = \hbar k \Gamma \rho_{ee}$ that is proportional to the probability of being in the excited state $\rho_{ee}$, where $k$ is the wavevector of the photon. For low velocities, $k \cdot v \ll \Gamma$, this force can be expanded \cite{Leibfried2003}  as,
\begin{equation}
F \simeq \hbar k \Gamma \left[\rho_{ee} + \frac{\delta \rho_{ee}}{\delta v}v\right]_{v=0} \equiv F_0(1+\kappa v)
\end{equation}
where the damping parameter $\kappa$ is responsible for the cooling force during Doppler cooling. For a high saturation parameter, the viscous force of the laser beam becomes negligible, $F_0 \kappa \to 0$, as the probability of being in the excited state is no longer velocity-dependent,  $\rho_{ee} \to 1/2$. For a trapped particle, the static force $F_0$ provides no net energy transfer averaged over the course of a trap cycle. However, we can use it to supply energy to the system if it is pulsed for a time faster than the lowest trap period.

To estimate the transferred energy during the pulse, we note that for an ion trapped in a potential $m \omega^2 x^2/2$, the presence of a constant force $F_0 = \hbar k_z \Gamma/2$ from the heating laser will displace the equilibrium position from $x =0$ to $x_{\rm{eq}} = F_0/m \omega^2$, approximately $46/\sqrt{2}$~nm for $\omega = 2 \pi \times 1$ MHz, where we assume $k_z=k/\sqrt{2}$, i.e. an angle of 45\textdegree between the laser beam and the axis of the mode. Therefore, turning on the heating laser for a duration of half the trap period will displace the ion by $2x_{\rm{eq}}$, in the process supplying the energy $E_0=m \omega^2 (2x_{\rm{eq}})^2/2 = (\hbar k \Gamma)^2/2m \omega^2$, approximately 8 vibrational quanta. This energy increase is much larger than the energy transferred due to momentum diffusion caused by the spontaneous emission and discreteness of the absorption process, which we estimate to be approximately $2$ motional quanta per half trap cycle \cite{Leibfried2003}. By repeating the laser pulses synchronized to the ion oscillation, increasingly more energy is supplied each time, and after $n$ pulses, the ion energy amounts to $E_n=n^2E_0$. 

We can study the equilibrium temperature distribution along the ion chain by focusing two laser beams individually onto the two ends of the ion chain. Each laser beam emulates a thermal bath for the outer ions. Measurement of the temperature of the ions along the chain can be done to examine the linear behavior expected for macroscopic systems. In the case of a trapped ion chain, the non-linear dynamics and quantum effects may lead to deviations from a linear temperature distribution across the chain \cite{Lin2010}. The temperature measurement may be performed by measuring the linewidth of the dipole transition or, if a reliable measure at lower temperatures is sought, via a side-band spectroscopy \cite{Leibfried2003,Haeffner2008}. 

\begin{figure}
\begin{center}
\includegraphics[width = 1\textwidth]{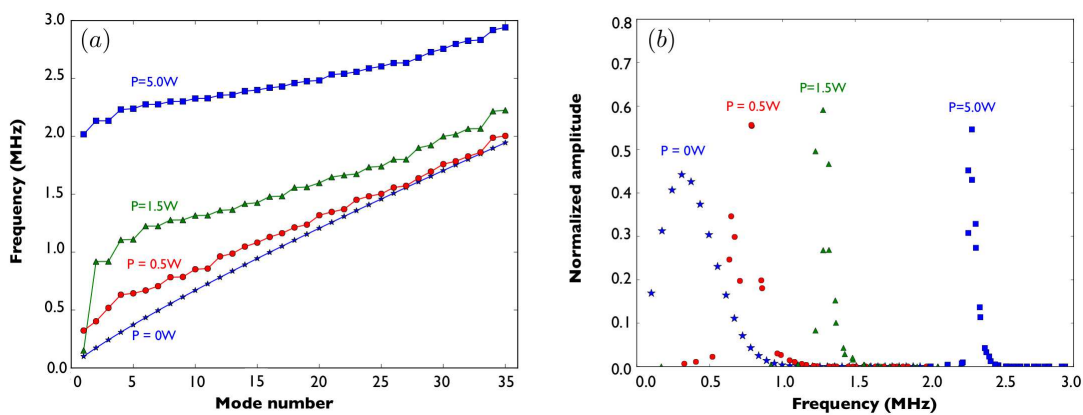}
\caption{\label{excitationspectrum} (a) The normal axial mode frequencies of a 35-ion string at different optical lattice powers. (b) The amplitude of each mode for a 35-ion string when a single ion at the end of the chain is excited axially.}
\end{center}
\end{figure}

We perform numerical calculations to establish the expected propagation time of the excitations by assuming that the ions undergo small displacements from their equilibrium positions along the axial direction. Then, we compute the normal mode eigenvectors, $\vec{v}_n$, and the corresponding eigenfrequencies, $\omega_n$, shown in figure~\ref{excitationspectrum}a, of the $N$-ion chain by linearizing the Hamiltonian \eref{FKH} around the ion equilibrium positions within the combined harmonic and optical potentials. We expand the excitation of a single ion at the end of the chain, $\hat{e_1} = [1,0,0,0,...]$, in terms of the eigenmodes of the ion chain,
\begin{equation}
\hat{e_1} = \sum_{n=1}^{N} c_n \vec{v}_n,
\end{equation}
where $c_n$ is the amplitude of the $n$-th normal mode. For no initial velocities, the time evolution of the last ion position $x_N(t)$ is given by
\begin{equation}
x_N(t) = \hat{e_N} \cdot \sum_{n=1}^{N} c_n \vec{v}_n \cos(\omega_n t).
\end{equation}

We find that even for long ion chains, the mode structure of the excitation $\hat{e_1}$ is dominated by the lower frequency modes as shown in figure~\ref{excitationspectrum}b. Due to the axial symmetry of the potential, each normal mode has a definite parity \cite{Morigi2004a}. In an even parity mode, the ions at the end of the chain oscillate in the same direction with the same magnitude, as, for instance, in the center of mass mode, and in an odd parity mode they oscillate in the opposite direction with the same magnitude, as in the case for the breathing mode. It follows that for even modes, the decomposition of the excitation of $\hat{e_N}$ is the same as the decomposition of $\hat{e_1}$, while for odd modes $c_{n} \to -c_{n}$ holds. Therefore, in order for the excitation to fully propagate to the other end of the ion chain, $x_N(t) = 1$,  the even parity modes have to evolve for an integer number of time periods and odd parity modes for a half integer such that $\cos(\omega^{\rm{even}}_n t) = +1$ and $\cos(\omega^{\rm{odd}}_n t) = -1$.

\begin{figure}
\begin{center}
\includegraphics[width = 1\textwidth]{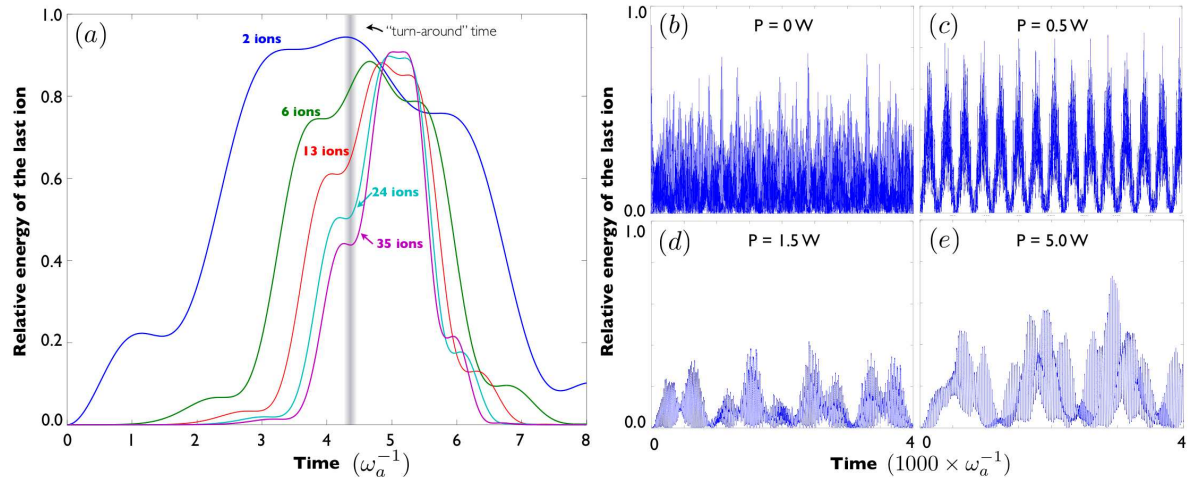}
\caption{\label{lastionpos} (a) The energy of the last ion relative to the energy of the initially displaced first ion without the optical potential as a function of time in units of $\omega_a^{-1}$. (b-e) The relative energy of the last ion of a 35-ion chain when the first ion is initially displaced from equilibrium for different lattice powers. Note that the time axis is in units of ($1000\times\omega_a^{-1}$).}
\end{center}
\end{figure}

However, from the calculations it follows that without the lattice, the last ion acquires a significant amount of energy long before all the modes acquire the appropriate phase. In fact, the energy of the $N$-th ion is significant after a ``turn-around" time, where the phase difference of the center of mass and breathing mode is $\pi$ (see figure~\ref{lastionpos}a). Here, we define the localized energy of the last ion to be $E^{\rm{loc}}_{n=N}(t) = \frac{1}{2} m \omega_{\rm{loc}}^2 x^2_{N}(t) + \frac{1}{2} m \dot{x}^2_{N}(t)$ with the local oscillation frequency $\omega_{\rm{loc}}^2 = (1/m)\partial^2 H/\partial x_N^2$. This measures the energy of the ion when other ions are fixed, ignoring the energy of their coupled motions. The accuracy of this approximation improves in the presence of the optical lattice as the ion excitation becomes more localized. 

In the presence of the lattice, the modes with the two lowest eigenfrequencies no longer determine the energy transport time. As seen in figure~\ref{excitationspectrum}b, the peak of the mode spectrum increases in frequency  and the spectrum becomes narrower with increasing optical lattice. In the limit of a very strong confinement from the optical lattice, all normal mode frequencies are nearly degenerate and given by the curvature of the sinusoidal potential. Thus, the energy transport time is increased in the presence of the lattice. This is depicted in figure~\ref{lastionpos}b-e. 

This calculation does not account for the anharmonicity of the optical potential and the Coulomb interaction. Furthermore, the depth of the optical lattice can only support few bound vibrational states, implying that a fully quantum mechanical analysis is necessary for a more complete understanding. For a purely sinusoidal potential produced by the lattice with power  1.5~W, we expect six bound states with the energy of the second level different by approximately 2 \% from the energy of a harmonic oscillator with the curvature at the minimum of the sinusoidal potential.

The ion-oscillator chains are an excellent system to compare basic thermodynamic properties such as the diffusion coefficient, heat capacity, latent heat, and coefficient of thermal expansion to \textit{ab initio} theories, both in the classical and the quantum regime. The experimental scheme presented above can be used for the study of latent heat and heat capacity of ion crystals \cite{Morigi2004a}, as measuring these thermodynamic properties involves adding heat into the system and then recording the temperature. The heat can be supplied via a laser or electronic noise. The control over the system's temperature provides a potential pathway to investigate the recent idea of a temperature-driven structural phase transition \cite{Gong2010}. Furthermore, a driven FK chain is a model frequently considered in the context of heat conduction in nanoscale devices. When one end of the lattice is driven periodically, the heat flux is calculated to exhibit a resonant behavior \cite{Ai2010}.  The presence of such a resonance could be probed with an ion string in an optical lattice. 

\section{Spin-dependent dynamics}\label{sec:spin-motion-dynamics}

In this section, we discuss experiments in which the dynamics of the ion crystal depend on the internal degree of freedom of the ions. This realises a spin-boson model where the spin degree of freedom interacts with the phonons of the ion crystal \cite{Porras2008}. Such spin-boson models have been extensively studied theoretically as they apply to the interaction of two-level systems with thermal baths, thereby providing a theory of the decoherence for a spin-1/2 system (see ~\cite{Leggett1987,Grifoni1998,Weiss1999}).

On the other hand, the coherence properties of the spin can be used to study the system playing the role of the thermal bath. For instance, one can detect the presence of chaos from the coherence decay of the spin \cite{Emerson2002}. This method has already been employed to detect signatures of quantum chaos \cite{Ryan2005} and even quantum critical points \cite{Zhang2008} in nuclear magnetic resonance experiments. Therefore, measuring the coherence of a spin connected to the system of interest is a valuable tool to characterise the phase space structure of complex quantum systems and detect quantum phase transitions.

\begin{figure}
\begin{center}
\includegraphics[width = 0.9\textwidth]{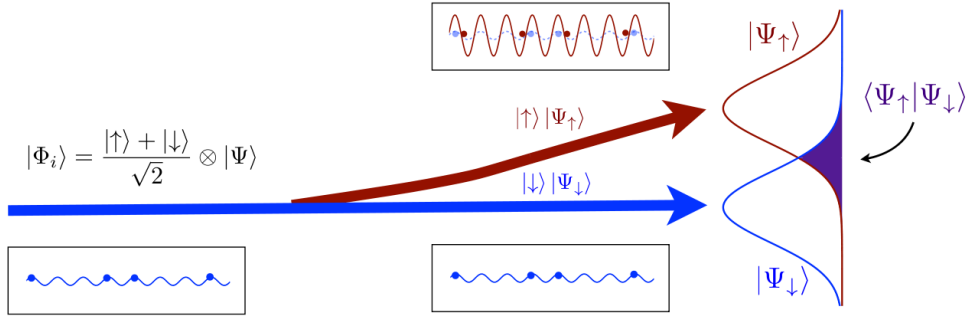}
\caption{\label{fig:interference} Interference experiment to determine the overlap between two quantum simulations. A spin is brought into a superposition of $\ket{\uparrow}+\ket{\downarrow}$. The wavefunction evolves under a spin-dependent Hamiltonian, resulting in slightly different dynamics for the two paths. Measuring the spin-coherence afterwards, one extracts the overlap between the two wavefunctions.}
\end{center}
\end{figure}

We consider a system initially prepared in a superposition spin state, $(\ket{\uparrow}+\ket{\downarrow})/\sqrt{2}$, propagating along two slightly different paths conditioned on the state of the spin \cite{Gorin2006}, as shown in figure~\ref{fig:interference}. Let the Hamiltonian of the total system be $H(t) = H_0(t)+ \ketbra{\uparrow}{\uparrow} \epsilon V(t)$. Here, $\epsilon V(t)$ is a small perturbation which acts only if the ion is in the state $\ket{\uparrow}$. We will show how measuring the coherence of the spin yields the overlap (or fidelity) between the two evolutions of the resulting wavefunctions \cite{Gorin2006,Gardiner1997}. Thus, by measuring the decoherence rate of the spin, we measure how strongly the perturbation $\epsilon V(t)$ leads to orthogonal wavefunctions, resulting in the rate at which the phase space mixes.

Let $\ket{\Psi}$ be the initial motional state of the ion string and $(\ket{\uparrow}+\ket{\downarrow})/\sqrt{2}$ be the spin state of one of the ions. The total initial wavefunction is $\ket{\Phi_i}=(\ket{\uparrow}+\ket{\downarrow})/\sqrt{2}\otimes\ket{\Psi}$. Applying the spin-dependent Hamiltonian, the wavefunction becomes
\begin{eqnarray}
\ket{\Phi'}&=&\frac{\ket{\uparrow}\ket{\Psi_\uparrow}+\ket{\downarrow}\ket{\Psi_\downarrow}}{\sqrt{2}},\:\end{eqnarray}
where $\ket{\Psi_\uparrow}$ and $\ket{\Psi_\downarrow}$ are the motional wavefunctions associated with spin $\ket{\uparrow}$ and $\ket{\downarrow}$, respectively.
We apply a $\pi/2$ pulse, $R(\pi/2,\varphi)$, with phase $\varphi$ and obtain:
\begin{eqnarray}
\ket{\Phi_f}&=&\frac{(\ket{\uparrow}+ie^{-i\varphi}\ket{\downarrow})\otimes\ket{\Psi_\uparrow}+(\ket{\downarrow}+ie^{i\varphi}\ket{\uparrow})\otimes\ket{\Psi_\downarrow}}{2}\:.\end{eqnarray}
Finally, we measure the probabilty to obtain spin up  $P_{\uparrow}$:
\begin{eqnarray}\label{eq:fidelity-decay-meas}
P_{\uparrow}(\varphi)&=& \left|\braket{\uparrow}{\Phi_f}\right|^2 \nonumber\\
&=&\frac{1}{4}\left| \ket{\Psi_\uparrow} + ie^{i\varphi}\ket{\Psi_\downarrow} \right|^2 \nonumber\\
&=&\frac{1}{2}(1+{\rm Re}({ie^{i\varphi}\braket{ \Psi_\uparrow}{\Psi_\downarrow}}))\:.
\end{eqnarray}
Measuring $P_{\uparrow}(\varphi)$ yields the scalar product $\braket{\Psi_\uparrow}{\Psi_\downarrow}$. Hence, with a simple single particle coherence measurement, we characterise the effect of the perturbation $\epsilon V(t)$ on an arbitrary complex quantum dynamics.

For experiments, a string of ions can be trapped in an optical lattice as discussed in the previous sections. The dynamics of the system under investigation can now be accessed by tuning the polarization of the trapping lattice to be elliptical. Thus, a \Ca ion in the $\ket{S_{1/2}, m_J=1/2} = \ket{\uparrow}$ state would experience a different force than one in $\ket{S_{1/2}, m_J=-1/2} = \ket{\downarrow}$, thus allowing the experimenter to control $\epsilon V(t)$.

We note that, by using this interferometric scheme, one determines the sensitivity of the quantum simulation to parameter fluctuations without the need for tedious quantum state tomography. Furthermore, the method allows us to test whether the system is near a quantum critical point \cite{Zhang2008}. In particular, it would be exciting to probe the Aubry transition, effectively preparing a superposition of two different quantum phases and comparing their time evolution directly with each other.

The spin-lattice interaction also can be used to imitate impurities in a solid. In such a scenario, the internal electronic states of individual ions would be prepared according to a random distribution. The state dependent interaction will then alter the local potential of these ions and, thus, emulate scatterers. It would be of particular interest to study the ground state properties as well as phonon propagation and dynamical Anderson localization. Finally, quantum control of the internal ion state allows for superpositions of potentials, leading to a new class of experiments with trapped ions.

\section{Conclusions}

We show that a trapped ion chain embedded in microtraps can be used to study oscillator models related to dry friction and energy transport. Classical numerical calculations assuming accessible experimental parameters demonstrate that an ion string trapped inside a short-wavelength lattice exhibits a structural phase transition similar to the Aubry transition in the Frenkel-Kontorova model. Furthermore, tuning the strength of the optical lattice changes the excitation spectrum of the ion chain, altering the energy transport efficiency. Finally, we present a scheme to characterise arbitrary complex quantum dynamics by measuring the coherence of a single spin. 

\ack
We thank Christopher Reilly for comments on the manuscript. This work was supported by the NSF CAREER program grant \# PHY 0955650. MR was supported by an award from the Department of Energy (DOE) Office of Science Graduate Fellowship Program (DOE SCGF). 

\section*{References}

\bibliographystyle{iopart-num}
\bibliography{../../../group/literature/bibtex/library}

\end{document}